\documentclass[aps,prb,twocolumn,nopacs]{revtex4}
\usepackage{graphicx}
\usepackage{times}
\usepackage{amsmath}
\usepackage{amsthm}
\usepackage{amssymb}
\usepackage{amsbsy}
\usepackage{amsfonts}
\newcommand{\abs}[1]{\left\vert #1 \right\vert}
\newcommand{\bra}[1]{\left\langle #1 \right\vert}
\newcommand{\ket}[1]{\left\vert #1 \right\rangle}

\begin{document}
\title{Improved phase gate reliability in systems with neutral Ising anyons}

\author{David J. Clarke}
\affiliation{Department of Physics and Astronomy, University of California,
Riverside, CA 92521, USA}
\author{Kirill Shtengel}
\affiliation{Department of Physics and Astronomy, University of California,
Riverside, CA 92521, USA}

\begin{abstract}
    Recent proposals using heterostructures of superconducting and either topologically insulating or semiconducting layers have been put forth as possible platforms for topological quantum computation. These systems are predicted to contain Ising anyons and share the feature of having only neutral edge excitations. In this note, we show that these proposals can be combined with the recently proposed ``sack geometry'' for implementation of a phase gate in order to conduct robust universal quantum computation. In addition, we propose a general method for adjusting edge tunneling rates in such systems, which is necessary for the control of interferometric devices. The error rate for the phase gate in neutral Ising systems is parametrically smaller than for a similar geometry in which the edge modes carry charge: it goes as $T^3$ rather than $T$ at low temperatures. At zero temperature, the phase variance becomes constant at long times rather than carrying a logarithmic divergence.
\end{abstract}
\maketitle

\label{sec:intro}
A device implementing topologically protected quantum computation would allow the fault tolerant manipulation of quantum information stored in the non-local properties of non-Abelian anyons.\cite{Kitaev03,Nayak08} Unfortunately, such a device is difficult to create, as the non-Abelian anyons that are most likely to be accessible in a laboratory setting, Ising anyons, are not associated with topologically protected operations that are universal for quantum computation. However, Ising anyons can store quantum information, and their manipulation can produce a protected set of operations (the Clifford gates) that are sufficient for any classical operation.\cite{Freedman02a,Freedman02b,Bravyi06} In order to make the jump to quantum computation, a single qubit phase gate (or a gate equivalent through Clifford operations) is necessary.\cite{Freedman06} Fortunately, by using a so-called ``magic-state distillation'' procedure we can tolerate a high degree of error (14\%) in the phase gate if the other basic operations are protected.\cite{Bravyi06}

Recently, we proposed a method for implementing a single-qubit phase gate in an Ising-anyon-based quantum computational platform.\cite{Bonderson10b} This was the first proposal that could reasonably meet the error threshold set forth by Brayvi~\cite{Bravyi06} in the context of this architecture. In principle, using this gate, a $\nu=5/2$ fractional quantum Hall (FQH) state could be made universal for quantum computation. However, building quantum devices in $\nu=5/2$ FQH systems presents several problems. First, the $\nu=5/2$ state itself is not fully understood, although recent evidence seems to point to its non-Abelian nature.\cite{Willett09a,Willett09b,Bishara09a} Second, high quality samples are necessary for the observation of the plateau at $\nu=5/2$, as well as ultra-low-temperature environments; the proposed phase gate, for instance, has an error rate that grows as $T$ at low temperatures. Finally, stray quasiparticles are likely to exist in this system even under ideal experimental conditions. In fact, their existence is the basis for the experiment purported to show the non-Abelian nature of excitations at $\nu=5/2$.\cite{Willett09a,Willett09b} In the context of quantum computing, these stray quasiparticles would interfere with the fine control of the system necessary to complete a full quantum algorithm.

New heterostructure systems based on $s-$wave superconductors (SC) sandwiched with either topological insulators (TI) or semiconductors with strong spin-orbit interactions have been proposed recently; the excitations in these systems are predicted to be Ising anyons.\cite{Fu08,Sau10a,Alicea10a,Lee09} These systems may have certain advantages compared to the other candidates for hosting non-Abelian excitations, in particular the $\nu=5/2$ fractional quantum Hall liquid, because they involve materials that may not have to meet requirements as stringent as those necessary for the FQHE. One characteristic of all these systems is that they have chiral edge modes. However, unlike in FQH systems, the edge modes with non-Abelian anyonic nature in these new systems are neutral. This presents a challenge because the interferometry experiments purported to show the non-Abelian nature of the anyonic excitations in the $\nu=5/2$ state\cite{Willett09a,Willett09b} rely on the electric signal due to their charge. However, recent theoretical developments have shown that, given the proper geometry, the interference of the neutral edge modes in the heterostructure systems may be electrically detected as well.\cite{Akhmerov09a,Fu09b,Nilsson10a,Sau10c} This makes relevant the question of the practicality of such systems for quantum information processing, and in particular warrants a reinvestigation of the phase gate for neutral systems. We find that in systems that have only neutral edge excitations the error rate of the phase gate actually improves significantly. However, these systems present a separate challenge in that the edge of the system is not so easily deformed as it is in a FQH system; our earlier proposal for the phase gate relied on being able to shape the edge by electrostatic gating in order to control tunneling rates. Here, we propose an alternative method of controlling such tunneling.

\begin{figure}[t]
\includegraphics[width=0.7\columnwidth]{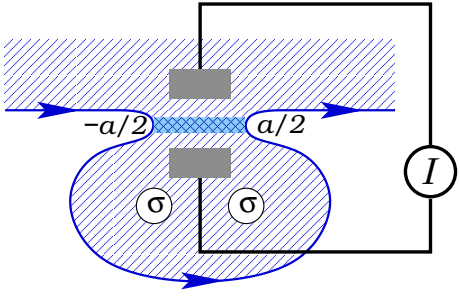}
   \caption{A top view of the proposed phase gate device in heterostructure systems with Ising anyons.
   A hatched region corresponds to the region coated with a superconductor; the edge state is formed at the boundary of this coating with a Zeeman-inducing magnetic material.  The SC-coated sack like region is pierced by two superconducting vortices, each binding a $\sigma$ anyon. A crosshatched region depicts a weak Josephson link, which is biased by the supercurrent $I$ in order to control the quantum phase slip rate (see text for details).}
\label{fig:gate}
\end{figure}

We shall begin by reviewing the analysis of Ref.~[\onlinecite{Bonderson10b}] in the context of a system with a single neutral edge mode.
Following that proposal, we consider a topological
qubit encoded in a pair of anyons, each carrying Ising topological charge
$\sigma$ whose two possible fusion
channels $I$ and $\psi$ form a computational basis.
In superconducting heterostructure systems,
the anyons comprising the qubits may be localized using superconducting vortices (or non-superconducting regions penetrated by half a quantum of fundamental magnetic flux $hc/e$).
These two anyons are placed in the ``sack'' geometry as shown in
Fig.~\ref{fig:gate}.\footnote{The other two anyons of the conventional four-anyon encoding are situated outside the sack and hence can be ignored in our analysis.}  The quasiparticles moving along the edge can tunnel from $-a/2$ to $a/2$ (so the sack has perimeter length $a$). Generally, there could be more than one type of excitations tunneling between $-a/2$ and $a/2$. In the weak-tunneling, low
temperature regime, the quasiparticles with the most relevant
tunneling operators will dominate the tunneling current. However, quasiparticles that do not carry $\sigma$ will have no effect on the topological qubit here and therefore will not enter in our analysis.

Let us pause here in order to clarify an important point which appears to be somewhat obscure in the context of superconducting heterostructure systems. The chiral Majorana fermions that are present on the edge of these systems are \emph{not} non-Abelian excitation, while Majorana zero modes bound to the vortices \emph{are}. In other words, it is the presence of vorticity, or a twist in the boundary conditions for the Majorana fermions, which is essential here. It is this twist that is encoded in the field $\sigma$. Therefore, the excitations that must tunnel in our device in order to effect the phase gate are \emph{vortices}. This point has been made in Refs.~[\onlinecite{Akhmerov09a,Nilsson10a}], albeit it was originally expected that these would be Abrikosov vortices. The serious drawback of Abrikosov vortices is that they tend to behave as very classical objects, which makes the prospects of their quantum tunneling or interference rather problematic. To the best of our knowledge, such quantum effect have not been observed in any systems. On the other hand, Josephson vortices can be quantum: their tunneling is known as quantum phase slips and their quantum interference has been observed in the context of Aharonov-Casher effect.\cite{Elion93,Elion94} Therefore the device should be fabricated with a Josephson weak link between $-a/2$ and $a/2$; we should rely on controlled tunneling of Josephson vortices along this link. We will revisit the issue of manipulating their tunneling amplitude later, for now we shall return to our analysis and show how the interference between the possible trajectories from left to right enacts a non-trivial transformation on the qubit.

 The combined edge and
qubit system is described by the Hamiltonian
\begin{equation}
H=H_E \otimes \openone + H_{\mathrm{tun}}(t) \otimes \boldsymbol{\sigma}_z,
\end{equation}
where $H_E$ is the Hamiltonian describing the unperturbed edge and
$H_{\mathrm{tun}}$ describes tunneling of Josephson vortices across
the constriction. Here the $\boldsymbol{\sigma}_z$ represents the braiding
statistics of the associated $\sigma$ with the qubit, picking up a minus sign
each time the $\sigma$ braids around the $\psi$ charge. In SC heterostructures this is the Aharonov-Casher phase induced by taking a vortex around a region with odd electron parity.~\cite{Hassler10a}

The density
matrix of the combined system is represented by $\chi$ and the qubit's density matrix is
obtained from this by tracing out the edge $\rho = \text{Tr}_{E} \chi$.
We
assume that the edge and the qubit are initially unentangled. Closely following Ref.~[\onlinecite{Bonderson10b}], we
solve the interaction picture Schr\"{o}dinger equation
\begin{equation}
\label{eq_density}
i \frac{\mathrm{d}\widetilde{\chi} (t) }{\mathrm{d}t}
= [\widetilde{H}_\mathrm{tun}(t)\otimes\boldsymbol{\sigma}_z,\widetilde{\chi}(t)],
\end{equation}
where $\widetilde{A}(t) = e^{i H_E t} A(t) e^{-i H_E t } $,  to obtain
\begin{equation}
\rho (t) =
\begin{bmatrix}
\rho_{00} (0)                        &     e^{-\varsigma^2/2} e^{-i \theta} \rho_{01} (0) \\
e^{-\varsigma^2/2}e^{i \theta} \rho_{10} (0)     &     \rho_{11} (0)
\end{bmatrix}.
\end{equation}
Here $\theta$ is the relative phase accumulated between the two basis states, while $\varsigma^2$ represents the loss of coherence due to variance in the phase. The diagonal elements of the qubit density matrix are unaltered from their initial state, as the Hamiltonian commutes with $\openone\otimes \sigma_z$.

Computing the values of $\theta$ and $\varsigma^2$ to second order in the tunneling Hamiltonian, we have
\begin{eqnarray}
\label{eq:theta_FQH}
\!\! \theta &\simeq&  2\int_{0}^{t}~\mathrm{d}t'
\left\langle \widetilde{H}_\mathrm{tun}(t') \right\rangle, \\
\label{eq:p_FQH}
\!\! \varsigma^2 &\simeq& -\theta^{2} + 4 \int_{0}^{t} \mathrm{d}t_1
\int_{0}^{t}~\mathrm{d}t_2 \left\langle \widetilde{H}_\mathrm{tun}(t_1)
\widetilde{H}_\mathrm{tun}(t_2) \right\rangle.
\end{eqnarray}

To compute concrete values of $\theta$ and $\varsigma^2$, we use the field
theoretic description of the edge of a system with a neutral Ising chiral edge mode. The Lagrangian for the unperturbed edge is
\begin{equation}
\label{lagrange}
L_E = i\int\!\mathrm{d}x \psi(\partial_t+v\partial_x )\psi,
\end{equation}
where $\psi$ describes the chiral fermion mode.
The operator that tunnels $\sigma$ quasiparticles
across the constriction is
\begin{equation}
H_{\mathrm{tun}} = \Gamma e^{-i\beta} \, \sigma \! \left(\mbox{$\frac{a}{2}$}\right) \sigma \!
\left( -\mbox{$\frac{a}{2}$} \right)  + h.c.,
\end{equation}
where $\beta$ includes the dynamical phase acquired in traveling around the sack as well as any Abelian braiding statistics factors.

Assuming the edge was initially in thermal equilibrium (i.e. $\chi(0) =
\frac{ e^{-H_{E}/T}}{ \text{Tr}_{E} \left[ e^{-H_{E}/T} \right] } \otimes
\rho(0)$) at temperature $T$, we find
\begin{equation}
\label{eq:gatefreq}
\left\langle \widetilde{H}_\mathrm{tun}(t) \right\rangle=2
\left(\frac{\lambda\pi T/ v}{\sinh{\frac{a\pi T}{ v}}}\right)^{ 1/8 }
\!\!\!\!\abs{\Gamma}\cos\varphi,
\end{equation}
Here $\lambda$ is a short range cutoff and
$\varphi = \arg \left\{ \Gamma \right\} - \beta$.

In order to estimate $\varsigma^2$, we assume $\Gamma$ is abruptly turned on to some
constant value for the duration of time from $0$ to $t$, and all other
quantities are held fixed. In this case, we have
\begin{equation}
    \varsigma^2=
    \omega^2\int_0^t\!\!\!\mathrm{d}t_1\!\!\!\int_0^t\!\!\!\mathrm{d}t_2~\eta(t_2-t_1),
\end{equation}
where $\omega=2\left\langle \widetilde{H}_\mathrm{tun} \right\rangle$ while the tunneling is on, and
\begin{equation}
    \eta(t)=
    \sqrt{\frac{\Upsilon(t)^{-1/4}+\Upsilon(t)^{1/4}}{2}}-1,
    \end{equation}
with
    \begin{equation}
    \Upsilon(t)=1- \frac{\sinh^2\left(\frac{ a\pi T}{ v}\right)}
    {\sinh^2\left((i\delta-t)\pi T\right)}.
    \end{equation}

Note that $\eta(T,t)\rightarrow 0$ exponentially for long times at a rate proportional to the temperature. Therefore, if the gate is performed slowly in comparison with the sack time $a/v$ then the variance grows linearly:
\begin{equation}\label{eq:varapprox}
    \varsigma^2 \simeq  t\int_{-\infty}^\infty\!\!\!\mathrm{d}t'~\omega^2\eta(t')
    -\int_{-\infty}^\infty\!\!\!\mathrm{d}t'~\abs{t'}\omega^2\eta(t'),
\end{equation}
More precisely, we have for $t>a/v$
\begin{eqnarray}
 \varsigma^2&=&\omega^2 \int_{-t}^t\!\!\!\mathrm{d}t'~\left(t-\abs{t'}\right)\eta(t'),\nonumber\\
 &=&\omega^2 \int_{-\infty}^\infty\!\!\!\mathrm{d}t'~\left(t-\abs{t'}\right)\eta(t')
 +2\omega^2\int_{t}^\infty\!\!\!\mathrm{d}t'~\left(t'-t\right)\eta(t')\nonumber\\
 &=&\lambda t+\varsigma_0^2+\Xi(t),
\end{eqnarray}
where $\Xi(t)=2\omega^2\int_{t}^\infty\!\!\!\mathrm{d}t'~\left(t'-t\right)\eta(t')$ tends to 0 at long times.

For $t\gg a/v$, we may approximate

\begin{equation}
\eta(t,T)\leq\eta(t,0)=\frac{a^4}{64v^4t^4}+\mathcal{O}\left(\frac{ a}{vt}\right)^6,
\end{equation}
so that
\begin{equation}
\Xi(t)\leq
\left(\frac{\omega a}{v}\right)^2\left[\frac{1}{3}\left(\frac{a}{8 vt}\right)^2+\mathcal{O}\left(\frac{a}{vt}\right)^4\right]
\end{equation}
This confirms that for long times $t\gg a/v$ the phase variance approaches the linear form (\ref{eq:varapprox}).  At 0 temperature, the approach is a power law going as $t^{-2}$. At finite temperature, the approach will be even faster, with the nonlinear term decreasing exponentially.

The long-time decay rate $\lambda$ when only the neutral mode is present is given by
\begin{equation}
    \lambda_n=\omega^2\int\mathrm{d}t'\eta(t').
\end{equation}
At $T=0$, this integral is exactly 0. This may be seen by noting that the integrand $\eta$ has no poles in the lower half of the complex $t$ plane when $T=0$, and falls off as $\abs{t}^{-4}$ for large $\abs{t}$. To calculate $\lambda$ for non-zero $T$, we first make a change of variables $u=\sinh(\pi t' T)$, so that in the lower half of the complex $u$ plane $\eta$ has only one branch cut running from $u=-i$ to $u=-i\infty$ and no poles. This allows us to change the integral from one running along the real axis to one running along the branch cut. If $T\ll v/\pi a$, the integral may be approximated as
\begin{eqnarray}
\lambda_n&=&\frac{\omega^2}{\pi T}\int_{1}^\infty\mathrm{d}y\frac{2}{64y^4\sqrt{y^2-1}}\left[\left(\frac{T}{T_n}\right)^4+\mathcal{O}\left(\frac{T}{ T_n}\right)^6\right]\nonumber\\
&=&\frac{\omega^2}{\pi T_n}\left[\frac{1}{48}\left(\frac{T}{T_n}\right)^3 +\mathcal{O}\left(\frac{T}{T_n}\right)^5\right],
\end{eqnarray}
Where $T_n=v/\pi a$ is the characteristic temperature scale for the neutral mode in the sack.
This may be compared with the formula for a general Ising edge with charge and neutral modes
\begin{equation}
\lambda=\lambda_n+\frac{2\omega^2}{\pi\widetilde{T}}\left[\frac{T}{\widetilde{T}}\tan^2\!\varphi
+\frac{1}{3}\left(\frac{T}{\widetilde{T}}\right)^3+\mathcal{O}\left(\frac{T}{\widetilde{T}}\right)^5\right],
\end{equation}
when $t\gg 1/\pi T\gg a/v_{c,n}$, and where $\widetilde{T}=\tilde{v}/\pi a$ (\mbox{$1/\tilde{v}^2=(g_n-1/8)/v_n^2+g_c/v_c^2$}), with $v_{c,n}$ and $g_{c,n}$ the velocities and scaling dimensions for the charge and neutral modes, respectively. This formula for $\lambda$ may be derived using the above method with the more general form of $\eta$ derived in Ref~[\onlinecite{Bonderson10b}]. Note that when only the Ising neutral mode is present, $g_n=1/8$ and $g_c=0$, so $\widetilde{T}=\infty$. This eliminates the first order term in $T$ and the explicit dependence on the Aharonov-Bohm phase $\varphi$.

We can see that having only a single Ising neutral mode provides a significant advantage over having charge and neutral modes both present. First, there is a marked decrease in the decay rate for neutral-mode-only systems, with $\lambda\sim T^3$ rather than $T$ at low temperatures. Second, the error does not depend explicitly on the Aharonov-Bohm phase, so there is no runaway error as the phase advancement slows near $\varphi=\pi/2$ as there is in the case where charge modes are present. Finally, the zero-temperature limit of the neutral system exhibits only a constant level of degradation in the qubit state at long times, while in general one would expect the off diagonal elements of the density matrix to decay as a power law. That is, in general $\Xi(t)+\varsigma_0^2$ has a logarithmic divergence for long times when $T=0$. In a neutral system, $\Xi(t)+\varsigma_0^2$ converges as $t\rightarrow\infty$.

The above analysis suggests that the most important temperature scale for a device intended to implement quantum operations via interferometry on a system with neutral Ising edge modes is $T_n=\hbar v/k_B a\pi$, where $a$ is the typical perimeter size of a device element such as a  sack\cite{Hou06,Bonderson10b} or an interferometer.\cite{Wan08,Bishara09a} The edge mode velocity $v$ is determined by the Fermi velocity in the host material (for devices based on topological insulators) or by the Rashba spin-orbit coupling constant (for semiconductor devices), as well as by how closely the chemical potential is tuned to the Dirac point.

Fu and Kane\cite{Fu09b} calculated the edge mode velocity in a SC/TI heterostructure to be
\begin{equation}
v=v_F\frac{\sqrt{1-\mu^2/M^2}}{1+\mu^2/\Delta^2},
\end{equation}
where $M$ is the energy of Zeeman splitting and $\Delta$ is the induced pairing potential due to the superconducting proximity effect.
In order to derive an upper bound on the operating temperatures of such systems, we assume that $\mu$ has been tuned to near zero and that the typical device perimeter $a$ is of order 1 $\mu$m. Then for the SC/TI system Be$_2$Se$_3$, which has Fermi velocity $4.6\times10^5$~m/s, the maximum operating temperature $T_n\lesssim 1$~{K}.
For the proposed SC/semiconductor system, we evaluate the edge mode group velocity given in Ref.~[\onlinecite{Sau10c}], i.e $v=\alpha\int\mathrm{d}x\bra{\phi}\sigma_x\tau_z\ket{\phi}$, where $\ket{\phi}$ is the Nambu spinor solution of the Bogoliubov-de~Gennes equations for the Majorana mode. For InAs, Ref.~[\onlinecite{Sau10a}] suggests the values of .5 meV for effective pairing potential and 1 meV for the Zeeman splitting  energy. Using these parameters and the measured Rashba coupling $1.4\times10^4$~{m/s} and effective mass $m^*\sim .05m_e$,\cite{Luo90} we have calculated the value of the integral to be .65. This leads to a velocity of $9\times10^3$~{m/s}. The maximum operating temperature is therefore $T_n\lesssim 20$~{mK}. In this sense the SC/TI system is the more promising proposal, though it may be possible to significantly increase the operating temperature in the semiconductor case by using a system with a giant Rashba coupling. (A Bi/Ag(001) interface, for instance, has been measured to have a Rashba coupling of $1.2\times10^6$~m/s.\cite{Nakagawa07} This corresponds to $T_n\lesssim 3$~{K}.)

\label{sec:control}
A new challenge presented by the proposed systems with neutral Ising anyons is that, unlike quantum Hall edges, their edges cannot be manipulated electrostatically by gating. E.g., in the case of TI based systems, the edge is defined by the boundary between two distinct coatings: those by a superconductor and a ferromagnet. It is extremely difficult to imagine how one would move such a boundary ``on the go'' in order to manipulate the tunneling rate of $\sigma$ particles (in this case, Josephson vortices) -- a key to implementing our proposal for a phase gate. A use of multiferroics might be of some utility here, yet, to the best of our knowledge, no device with a tuneable FM-SC boundary has been fabricated to the date. What we envision instead is using a relatively wide Josephson junction (to suppress the ``ambient'' vortex tunneling) and biasing it by current in order to manipulate the shape of the tunneling barrier. Indeed, the quantum dynamics of a Josephson junction is usually modeled in terms of a motion of a particle in a ``washboard'' potential $U(\phi)= -U_0(2 I \phi/I_{c} + \cos{2\phi })$ where $U_0=\hbar I_{c}/2e$ with $I_{c}$ being the critical current. The current across the junction provides the tilt which in turn makes the potential barrier for a phase slip both lower and narrower. In the language of Josephson vortices the same phenomenon is described in terms of a Magnus force acting on a vortex due to the transverse current; the effect of such a force is to tilt the potential barrier for the vortex. Since the tunneling rate depends exponentially on the parameters of this barrier, current biasing should provide a simple and reliable way of manipulating such a rate for the purpose of effecting the phase gate.

\label{sec:conclusion}
To conclude, we have discussed both advantages and disadvantages arising from using the sack geometry for effecting a phase gate in the systems with neutral Ising anyons such as recently proposed heterostructure systems where either a topological insulator or a semiconductor with a strong spin-orbit coupling (and an additional source of Zeeman splitting) is interfaced with a superconductor. One significant advantage is a parametrically lower error rate while the most apparent disadvantage is a lack of developed techniques of manipulating the shape of the edges. The latter was the key to affecting the tunneling rates across the constriction in the original proposal. In this note we have proposed an alternative way of manipulating the tunneling rate using a weak Josephson link biased by current.

We should also note that while this paper was in preparation, another proposal for enacting a phase gate in this type of systems has been put forward~\cite{Hassler10a}. This proposal, while also relying on tunneling of Josephson vortices, uses a different idea: namely coupling a topological qubit to a flux qubit, and effecting a phase gate on the flux qubit instead.

\emph{Acknowledgments --}
It is our pleasure to acknowledge discussions with C.~W.~J.~Beenakker, P.~Bonderson, L.~I.~Glazman, P.~M.~Goldbart, F.~Hassler, C.-Y.~Hou, C.~Nayak, Yu.~V.~Nazarov and J.~D.~Sau. We would also like to acknowledge the hospitality of Microsoft Station Q, the Max Planck Institute for the Physics of Complex Systems and the Aspen Center for Physics, where many of these valuable discussions took place. DC and KS are supported in part by the DARPA-QuEST program, KS is supported in part by the NSF under grant DMR-0748925.

\bibliographystyle{apsrev}
\bibliography{../bibs/corr}
\end{document}